\documentclass[
	aps,
	amsmath,
	amssymb,
	reprint,
	superscriptaddress]{revtex4-2}

\usepackage{
	graphicx,
	dcolumn,
	bm,
	mathptmx,
	hyperref,
	soul,
	textgreek}
	
\usepackage[dvipsnames]{xcolor}

\DeclareUnicodeCharacter{2212}{-}

\hypersetup{
    colorlinks=true,
    linkcolor=[rgb]{0, 0.52, 0.73},
    filecolor=[rgb]{0, 0.52, 0.73},      
    urlcolor=[rgb]{0, 0.52, 0.73},
    citecolor=[rgb]{0, 0.52, 0.73},
}

\begin{document}


\title[Nonstanding spin waves in a single rectangular permalloy microstrip under uniform magnetic excitation]{Nonstanding spin waves in a single rectangular permalloy microstrip under uniform magnetic excitation}

\author{Santa Pile}
\affiliation{Institute of Semiconductor and Solid State Physics, Johannes Kepler University Linz, 4040 Linz, Austria}

\author{Sven Stienen}
\author{Kilian Lenz}
\author{Ryszard Narkowicz}
\affiliation{Helmholtz-Zentrum Dresden-Rossendorf, Institute of Ion Beam Physics and Materials Research, 01328 Dresden, Germany}

\author{Sebastian Wintz}
\author{Johannes F{\"o}rster}
\affiliation{Max Planck Institute for Intelligent Systems, 70569 Stuttgart, Germany}

\author{Sina Mayr}
\affiliation{Paul Scherrer Institut, 5232 Villigen PSI, Switzerland}
\affiliation{Laboratory for Mesoscopic Systems, Department of Materials, ETH Zurich, 8093 Zurich, Switzerland}

\author{Martin Buchner}
\affiliation{Institute of Semiconductor and Solid State Physics, Johannes Kepler University Linz, 4040 Linz, Austria}

\author{Markus Weigand}
\affiliation{Helmholtz-Zentrum Berlin f{\"u}r Materialien und Energie, 12489 Berlin, Germany}

\author{Verena Ney}
\affiliation{Institute of Semiconductor and Solid State Physics, Johannes Kepler University Linz, 4040 Linz, Austria}

\author{J{\"u}rgen Lindner}
\affiliation{Helmholtz-Zentrum Dresden-Rossendorf, Institute of Ion Beam Physics and Materials Research, 01328 Dresden, Germany}

\author{Andreas Ney}
\affiliation{Institute of Semiconductor and Solid State Physics, Johannes Kepler University Linz, 4040 Linz, Austria}

\date{\today}

\begin{abstract}
Ferromagnetic resonance modes in a single rectangular Ni\textsubscript{80}Fe\textsubscript{20} microstrip were directly imaged using time-resolved scanning transmission x-ray microscopy combined with a phase-locked ferromagnetic resonance excitation scheme and the findings were corroborated by micromagnetic simulations. Although under uniform excitation in a single confined microstructure typically standing spin waves are expected, all imaged spin waves showed a nonstanding character both, at and off resonance, the latter being additionally detected with microantenna-based ferromagnetic resonance. The effect of the edge quality on the spin waves was observed in micromagnetic simulations.
\end{abstract}

\maketitle

\section{\label{sec:introduction}Introduction}
Over the past decades charge-based computing devices decreased drastically in size, coinciding with an increasing and even limiting heat dissipation, which triggered an active search for new ways of information processing. In that regard, magnons or spin waves are to be one of the options to replace the transfer of electronic charges in logic devices \cite{002,003,004,039}. Therefore, the corresponding field of magnon spintronics nowadays attracts increasing attention as a promising direction of research \cite{001,005,006,007}. A wide range of geometrical magnonic systems of various materials has been investigated at this point, including thin films \cite{008,009,010}, multilayer nanostructures \cite{011}, magnonic crystals \cite{012,013,014,015}, magnonic waveguides \cite{016,017,018} and confined microstructures \cite{019,020,038}.

In this work the focus is put on a fundamental understanding of the dynamic magnetic properties of confined structures, as this is a prerquisite for the development of nanoscale computational devices. It was observed experimentally that the spins near the edges of a confined microstructure often behave as if they are ``pinned'' and, thus, are hindered to precess \cite{018}. Reflections at the edges cause spin-wave resonances, whenever the distance between the edges equals an integer number of half wavelengths, in other words, quantization of the spin-wave modes occurs \cite{006}. The spin-wave spectrum of a saturated ellipsoidal magnetic element can be calculated analytically \cite{034a,034b}. However, in most cases the magnetic elements considered for applications have a nonellipsoidal shape and are not saturated. It has been shown that both the nonellipsoidal shape and quality of the edges of these elements drastically affect their dynamic magnetic properties \cite{020,038}. Often spin waves are investigated using a nonuniform excitation of the structure \cite{002,004,010,012}. However, in micro- or nanoscale devices a uniform or close to uniform excitation field can be easier realized, e.g., placing an element in close proximity to an antenna. In a rectangular microstrip the pinning effect can be exploited to excite spin waves with an odd number of half wavelengths within the length of the strip using uniform radio-frequency\slash microwave (MW) driving fields \cite{006,031a}. The quantization conditions for the modes of a rectangular confined structure were described in detail in \cite{019}. Due to the high inhomogeneity of the effective field inside the structure along the external static magnetic field, the quantization conditions for a $k$-vector aligned in the direction of the external field are complicated \cite{048}. Therefore, analytical calculations of the spin-wave dispersion and consequent analysis are complex \cite{131}. A different approach that can be used for the spin-wave analysis of such structures is an experimental study in combination with micromagnetic simulations.

Experimental investigations on a micron to sub-micron scale require sensitive experimental techniques. The possibility to excite and image spin waves in micron-sized structures was shown to be possible using techniques such as Brillouin light scattering \cite{020,021}, time-resolved Kerr-microscopy \cite{056} or spatially resolved ferromagnetic resonance force microscopy \cite{022}. On the other hand, time-resolved scanning transmission x-ray microscopy (TR-STXM) \cite{027,027a,027b,027c,027d,027e} can be applied using a phase-locked ferromagnetic resonance (FMR) excitation scheme (STXM-FMR) \cite{030,031,032,033,038,073}. This STXM-FMR technique enables direct, time-dependent imaging of the spatial distribution of the precessing magnetization across the sample during FMR excitation with elemental selectivity.

The development of planar microresonators\slash microantennas allows for measuring FMR of a single ferromagnetic microstrip including resonance lines corresponding to spin-wave excitations \cite{023,024,025,026,032}. The presence of spin waves in a single Ni\textsubscript{80}Fe\textsubscript{20} (permalloy, Py) microstrip under uniform microwave excitation was evidenced earlier using microresonator-based FMR and supported by micromagnetic simulations \cite{025}. Those findings were corroborated by detailed investigations of comparable spin-wave excitations in a Co microstrip with similar dimensions \cite{026}. The spin waves in a confined microstructure, which are detectable with FMR, were expected to have a standing character \cite{034}, meaning that their amplitude minima should have remained at the same position in space over time. However, a nonstanding behavior of such spin waves was observed using STXM-FMR in a system of two rectangular Py microstrips arranged perpendicular to each other with a distance of 2\,\textmu m between them \cite{038}. The reason for the nonstanding character of the spin waves in one of the strips, as was reported in \cite{038}, may be inhomogeneities produced by the internal magnetization landscape of the strip and\slash or inhomogeneous external magnetic stray fields, emerging from the second strip. Both can lead to a superposition of spin-wave eigenmodes possibly resulting in the observed motion of the amplitude minima. In this work it is shown that nonstanding spin waves can be excited and directly imaged using STXM-FMR in a single Py microstrip. Furthermore, the spin waves, that exhibit a nonstanding character, can be detected with conventional microantenna FMR.

\section{\label{sec:expDet}Experimental Details}
\begin{figure}
	\centering
    \includegraphics[width=0.48\textwidth]{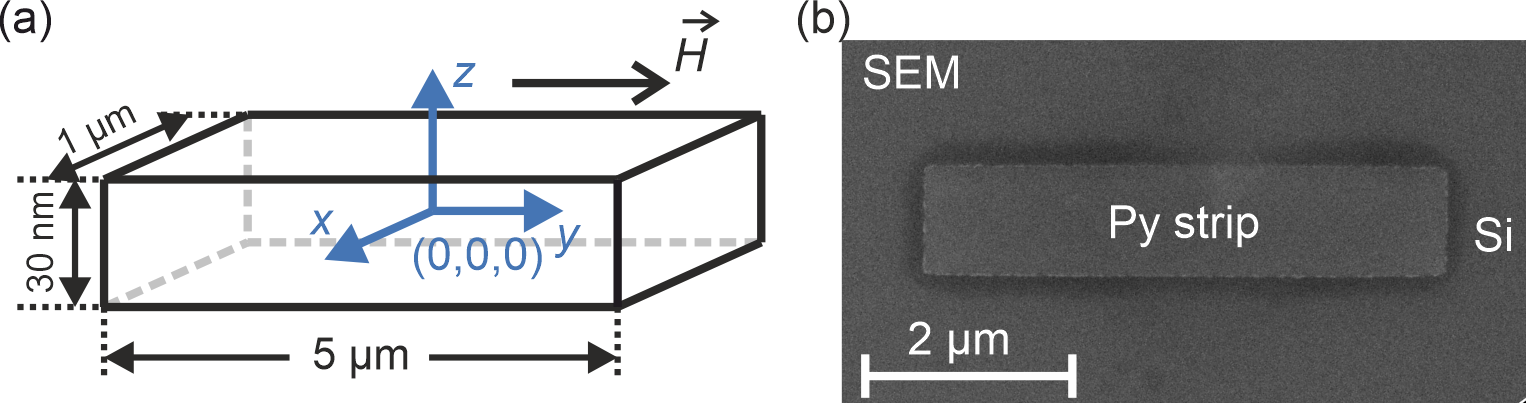}
    \caption{(a) Schematic representation of the rectangular magnetic element with indication of the directions of the coordinate axes, an example of in-plane external magnetic field $\vec{H}$ orientation and dimensions of the strip. (b) SEM image of the lithographically patterned Py rectangular microstrip.}
   \label{fig:stripDem}
\end{figure}

\begin{figure}
	\centering
    \includegraphics[width=0.35\textwidth]{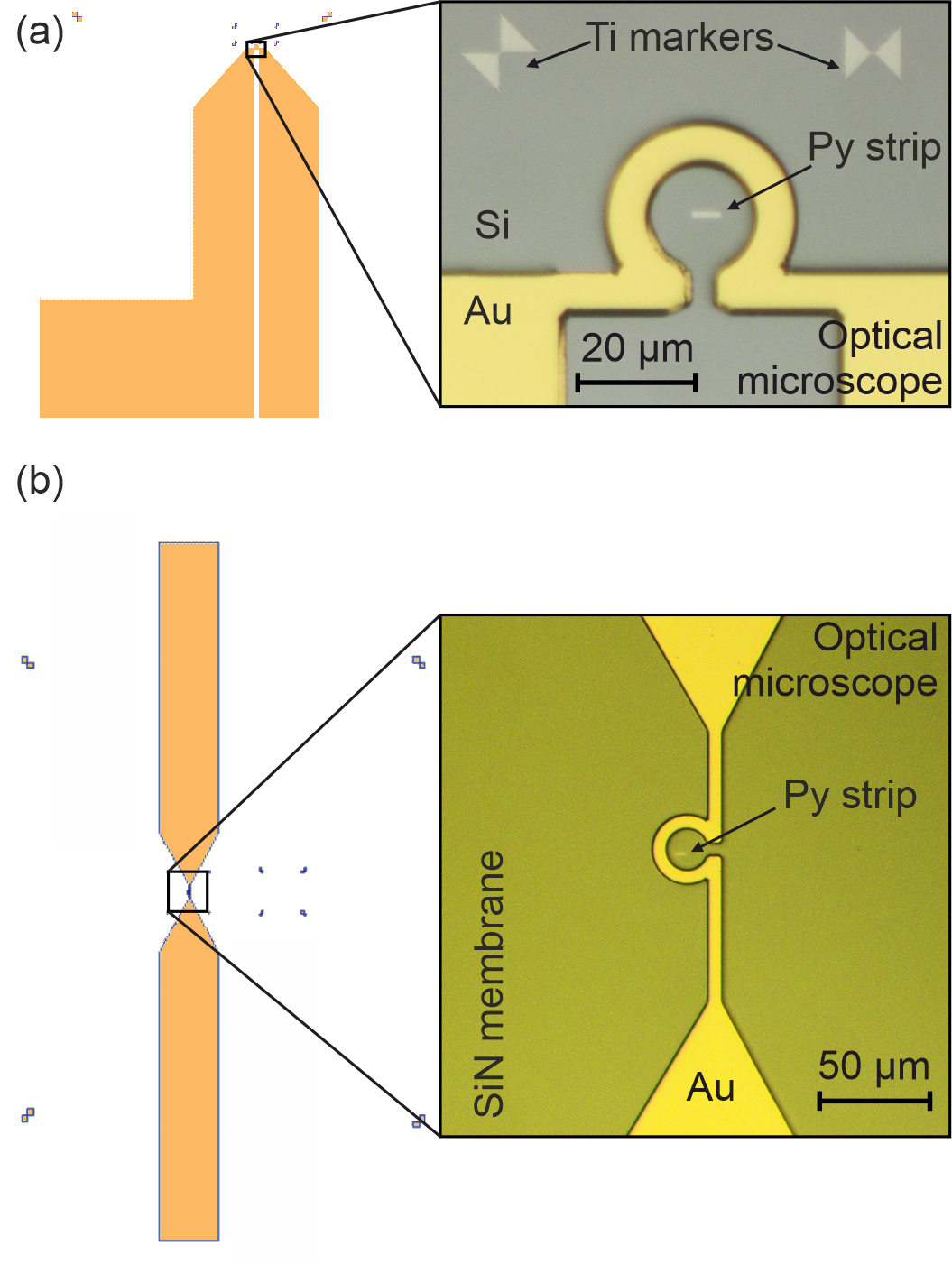}
    \caption{Schematics of the planar microantenna designs with close-up optical images of the loops: (a) multi-frequency microantenna, (b) microantenna for transmission excitation.}
   \label{fig:antennas}
\end{figure}

Single rectangular Py strips with a nominal lateral dimension of $5\times 1$\,\textmu m\textsuperscript{2} and 30\,nm thickness [see Fig.\,\ref{fig:stripDem}\,(a)] were fabricated on highly insulating Si substrates for the microantenna-based FMR measurements and on commercial 200-nm-thick $0.25\times 0.25$\,mm\textsuperscript{2} SiN-membranes for the STXM-FMR measurements using standard e-beam lithography and magnetron sputtering with subsequent lift-off. Sputtering was carried out at a process pressure of $4\times 10^{−3}$\,mbar in an ultrahigh vacuum chamber with a base pressure of $2\times 10^{−9}$\,mbar. Py with a nominal thickness of 30\,nm was sputtered at room temperature using 10 standard cubic centimeters per minute of Ar as process gas. To prevent oxidation the Py layer was covered with a 5\,nm thick Al capping layer grown using pulsed laser deposition. An example of the fabricated Py rectangular microstrip is shown in Fig.\,\ref{fig:stripDem}\,(b) imaged by a scanning electron microscope. In a second step, a planar microantenna \cite{033} with a Au thickness of 600\,nm was fabricated by photolithography, electron beam physical vapor deposition and a lift-off. In order to improve an adhesion of the Au layer, a 5\,nm Ti layer was placed between the Au and a substrate in the process. The two different designs of microantennas used in this study are shown in Fig.\,\ref{fig:antennas}. Design (a) was used for the microantenna-based FMR and (b) for STXM-FMR measurements.

For the microantenna-based FMR measurements a home-built MW spectrometer with field modulation at 78\,kHz and lock-in technique was used \cite{096,097}. The spectrometer was operating in field-sweep mode. The sample was placed in the microloop of a planar broadband microantenna with an inner diameter of 20\,\textmu m as can be seen in the optical microscope image in Fig.\,\ref{fig:antennas}\,(a). The microloop allows for the concentration of the MW field at the sample area \cite{023,024,025,032}, thus, enhancing the sensitivity of the system compared to conventional MW resonators. The MW field component was oriented perpendicular to the sample plane. The spectrometer operates in a Mach-Zender type interferometer scheme \cite{130} covering three frequency bands: 2--4\,GHz, 4--8\,GHz and 8--18\,GHz. The FMR absorption and dispersion signals were detected and rectified using homodyne mixing and fed into lock-in amplifiers working at the field-modulation frequency. For the measurements the external static magnetic field $\vec{H}$ was applied in the plane of the microstrip.

The STXM-FMR measurements were carried out at the MAXYMUS endstation of the UE46 undulator beam-line at the Helmholtz-Zentrum Berlin during the low-alpha operation mode of the BESSY II synchrotron. In STXM the x-ray beam is focused using a diffractive zone plate in combination with an order sorting aperture to remove perturbing undiffracted light and diffraction orders $\geqslant2$. The images are created by scanning the sample through the focused x-ray beam and detecting the respective x-ray transmission. For the measurements presented here the sample was scanned in steps of 50\,nm. The STXM-FMR technique exploits the x-ray magnetic circular dichroism effect as magnetic contrast mechanism. The latter allows for probing the dynamic out-of-plane magnetization component across the area of the microstrip when the circularly polarized x-rays are directed perpendicular to the sample surface. The photon energy was tuned to the Fe L$_3$-edge ($\sim 708$\,eV).

While scanning the sample through the focused x-rays, the magnetization dynamics were excited by applying a static magnetic field (in the range of $32$--$260$\,mT) and a small MW field ($\sim$\,$0.5$\,mT) in the same geometry as for the microantenna-based FMR measurements but using the microantenna design shown in Fig.\,\ref{fig:antennas}\,(b). The time resolution to probe the precessing magnetization at several intermediate points of its period can be obtained by a pump-and-probe measurement scheme, meaning that the MW frequency $f_\mathrm{MW}$ can be chosen depending on the frequency $f_\mathrm{s}$ of the x-ray bunches impinging on the sample (the so-called ``ring frequency'') \cite{031}:
\begin{equation}\label{eq:ET_freq}
f_\mathrm{MW}=\frac{f_\mathrm{s}}{N}\cdot M\,,
\end{equation}
\noindent where $N$ is the number of time channels or, in other words, the number of points over the excitation period at which the sample will be probed and $M$ is the number of excitation periods over the observation period ($N$ time spacings between x-ray pulses). In order to be able to probe the excitation period at $N$ different phases, $M$ and $N$ should not have common factors. The reported results were obtained at the excitation frequency of 9.43\,GHz ($f_\mathrm{s}=500$\,MHz, $M=132$, $N=7$). The magnetic contrast was extracted by dividing the counted x-ray photon signal at each time point by the time averaged value \cite{030,031,038,073}.

\section{\label{sec:results}Results}
\subsection{Micromagnetic Simulations}
\begin{figure*}
\includegraphics[width=0.8\textwidth]{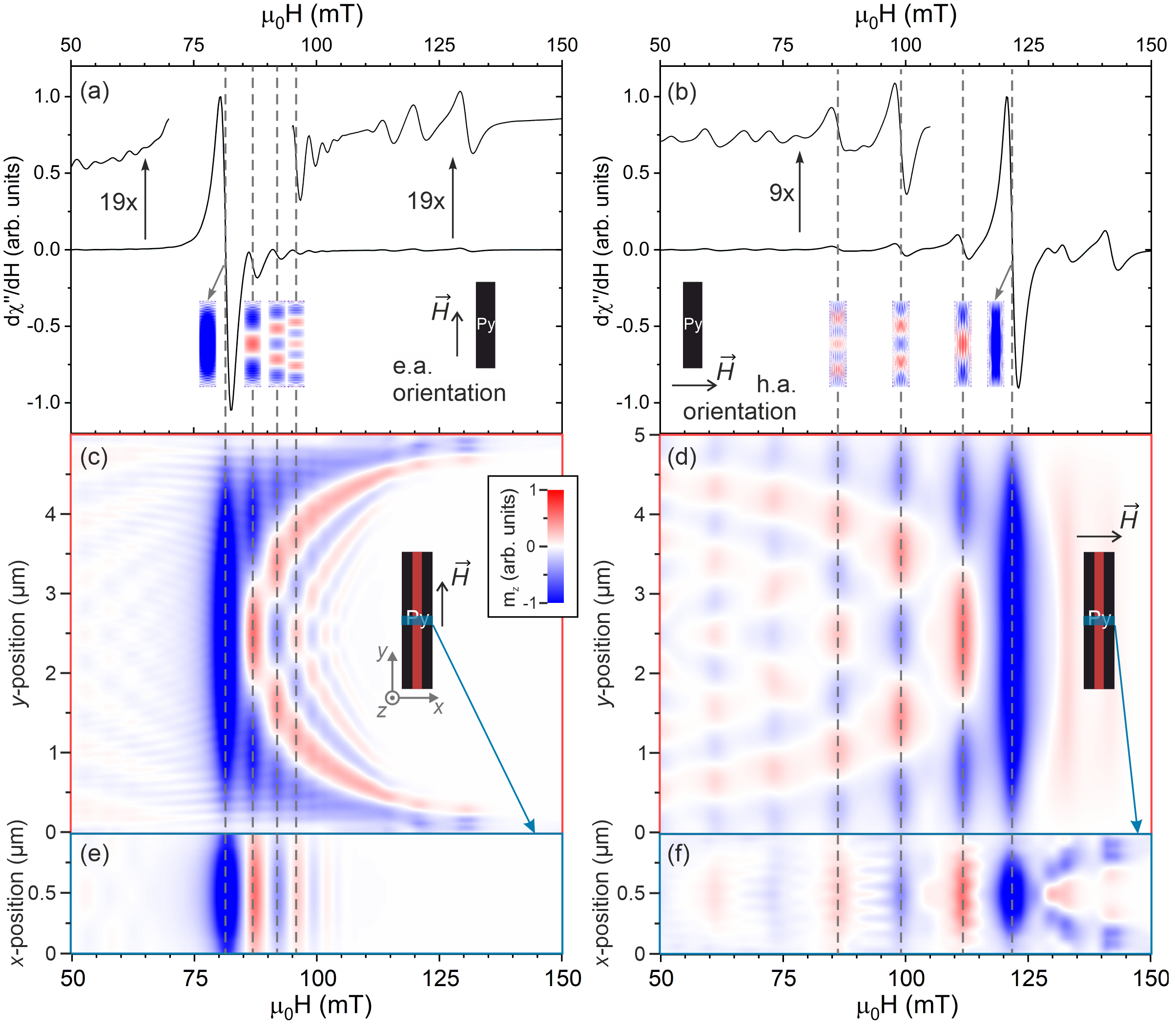}
\caption{(a,b) Simulated FMR spectra of the rectangular Py strip. (c-f) Simulated FMR spectra of the Py strip at 9.43\,GHz. Overview of the spin-wave profiles (c,d) along and (e,f) across the strip in (a,c,e) easy and (b,d,f) h.a.\,orientations.}
\label{fig:initsims}
\end{figure*}

Along with the experiment, simulations of the FMR spectra and the spatial distribution of the dynamic magnetization of the samples were carried out using \textsc{MuMax3} \cite{035}. In a first step the simulations were run for a rectangular strip [see Fig.\,\ref{fig:stripDem}\,(a)] with lateral dimension of $5\times 1$\,\textmu m\textsuperscript{2} and a thickness of 30\,nm. The cell size was chosen to be $\sim 10 \times 10 \times (sample\;thickness)$\;nm$^3$, which was sufficient for the dipolar-dominated spin waves imaged in this work. A saturation magnetization of $M_\mathrm{s}=700$\,kA/m \cite{053}, a Gilbert damping parameter $\alpha=0.006$, and an exchange stiffness constant of $A_\mathrm{ex}=13\times10^{-12}$\,J/m were used for the simulations. The magnetocrystalline anisotropy was set to zero. The frequency of the uniform MW field aligned along the z-axis [see Fig.\,\ref{fig:stripDem}\,(a)] was set to $f_\mathrm{MW}=9.43$\,GHz, which was used previously for the STXM-FMR measurements of similar samples \cite{033,038,073}, and its amplitude was set to 0.5\,mT. The resulting simulated FMR spectra for the in-plane field being oriented parallel to the long edge of the strip (easy axis orientation, e.a.\,orientation) and parallel to the short edge of the strip (hard axis orientation, h.a.\,orientation) are shown in Figs.\,\ref{fig:initsims}\,(a) and (b), respectively. The simulated FMR spectrum was obtained by derivation and normalization of the spectra of the out-of-plane magnetization component integrated over the strip area. The direction of the external magnetic field in both orientations is indicated in the insets of the figure. The magnitude of the external magnetic field was varied from 50\,mT to 150\,mT in 0.25\,mT steps. At each field the simulation was run for 50 excitation cycles in order to reach a steady magnetization precession state. The magnetization snapshots for each field were saved at the end of the 50\textsuperscript{th} excitation cycle, where the excitation field amplitude is 0 and, therefore, the resonance response shifted by $\frac{\pi}{2}$ with respect to the excitation field is maximum. The dashed vertical lines indicate some of the resonance field values. Blue-white-red images below the spectra show the calculated characteristic snapshots of the spatial distribution of the out-of-plane magnetization component $m_\mathrm{z}(t)$ at resonance. The value of $m_\mathrm{z}(t)$ changes across the area of the strip due to the variation of the  precession angle and the phase variation of the magnetization \cite{031a}. At the amplitude minima of the wave the precession angle approaches zero, consequently, $m_\mathrm{z}(t)\sim 0$, which is represented by the white color in the blue-white-red color scale provided in Fig.\,\ref{fig:initsims}\,(c).

In order to make an overview of the simulated $m_\mathrm{z}(t)$ profiles along the length and the width of the strip over the range of external field values at the fixed MW frequency, the averaged middle part of each spatial map was stacked into color plots as shown in Figs.\,\ref{fig:initsims}\,(c-f). In the figures the overviews of the spin-wave profiles along the strip (c,d) as well as across the strip (e,f) are shown for e.a.\ orientation (c,e) and h.a.\ orientation (d,f), respectively. The $x$-axis of the plots corresponds to the external magnetic field in mT. The $y$-axis shows either the position along ($y$-position) or across ($x$-position) the strip in \textmu m. The color and its intensity denote $m_\mathrm{z}(t)$ of the spin-wave profile on the same scale as in the blue-white-red images in Figs.\,\ref{fig:initsims}\,(a,b). The red and the blue rectangles placed over the Py strip schematic representation in Figs.\,\ref{fig:initsims}\,(c,d) indicate the regions of the Py strip that were used for averaging the data at each field for the color plots along and across the strip, respectively. The spatial distribution variation of $m_\mathrm{z}(t)$ in the directions parallel and perpendicular to the external field includes quantized k-vectors of the spin-wave eigenmodes excited in those directions. Therefore, the color plots allow for visual evaluation of the spin-wave eigenmodes, parallel and perpendicular to the strip, forming the resulting spin-wave pattern at each field. As a result, the overview visualization together with the FMR spectra as plotted in Figs.\,\ref{fig:initsims}\,(a-f) allows for analyzing the  excitable spin waves in the strip and helps revealing the relation between the FMR spectrum and the spin-wave excitations along and across the strip as a function of the external field.

In Figs.\,\ref{fig:initsims} (a,b) for each orientation of the strip one can observe the main FMR line with the largest intensity and several smaller signals, some of which are magnified in the insets of the plots. The spatial distributions of the out-of-plane dynamic magnetization component $m_\mathrm{z}(t)$ shown in the plots confirm that the main FMR signal in both orientations is the quasi-uniform excitation of the strip and smaller signals correspond to spin waves \cite{025,026}. The quasi-uniform excitation is the one, at which almost all magnetic moments across the strip area precess in phase. A nonuniform $m_\mathrm{z}(t)$ spatial distribution at the edges arises from the inhomogeneity of the effective field within the strip \cite{019,128}. When comparing the spectra in Figs.\,\ref{fig:initsims}\,(a) and (b) it is visible that due to the shape anisotropy the main FMR signal in e.a.\,orientation appears at lower fields (81.7\,mT) than in the h.a.\,orientation (121.7\,mT).

The overview analysis of the simulated spin-wave profiles of a perfectly rectangular Py microstrip plotted in Figs.\,\ref{fig:initsims}\,(c-f) shows that a rich spectrum of spin waves with an odd number of amplitude maxima can be excited using a uniform MW field in both considered orientations of the strip, e.a.\,and h.a. This observation is in agreement with previous research on the topic \cite{025,026}. Moreover, resonance field differences between neighboring odd spin-wave orders in the e.a.\,orientation [see Figs.\,\ref{fig:initsims}\,(a,c,e)] are much smaller than in the h.a.\,orientation [see Figs.\,\ref{fig:initsims}\,(b,d,f)]. For example, in e.a.\,orientation the resonance field difference between the waves with 3 and 5 amplitude maxima along the strip [Fig.\,\ref{fig:initsims}\,(c)] is 4.8\,mT, whereas in the h.a.\,orientation it is 12.6\,mT [Fig.\,\ref{fig:initsims}\,(d)]. The difference in the spin-wave mode separation along the strip for the two different field geometries can be attributed to the underlying fundamental dispersion relations: Backward-volume (BV) modes in e.a.\,orientation and Damon-Eshbach (DE) modes in h.a.\,orientation. For the e.a.\,orientation, the spin-wave modes are occurring at higher fields than the main FMR mode and their field-separation is relatively small. This corresponds to the magnetostatic BV dispersion relation lying below the FMR frequency and exhibiting a relatively high slope ($k$ changes strongly with frequency and, thus, field) \cite{006}. For the h.a.\,orientation, the spin-wave modes show up at fields below the main FMR mode and their field separation is relatively wide. This corresponds to the magnetostatic DE dispersion over frequencies above the FMR and showing a relatively high slope (k changes weakly with frequency and, thus, field) \cite{103}.

The FMR simulations were run and further adjusted in order to match the FMR measurements as will be discussed in section ''FMR Measurements``. In order to match the STXM-FMR measurements time-resolved micromagnetic simulations were performed providing information on the $m_\mathrm{z}(t)$ evolution over the time at each of the measured field values as will be discussed further in section ''STXM-FMR Imaging``.

\subsection{\label{sec:fmr}FMR Measurements}
\begin{figure}
\includegraphics[width=0.45\textwidth]{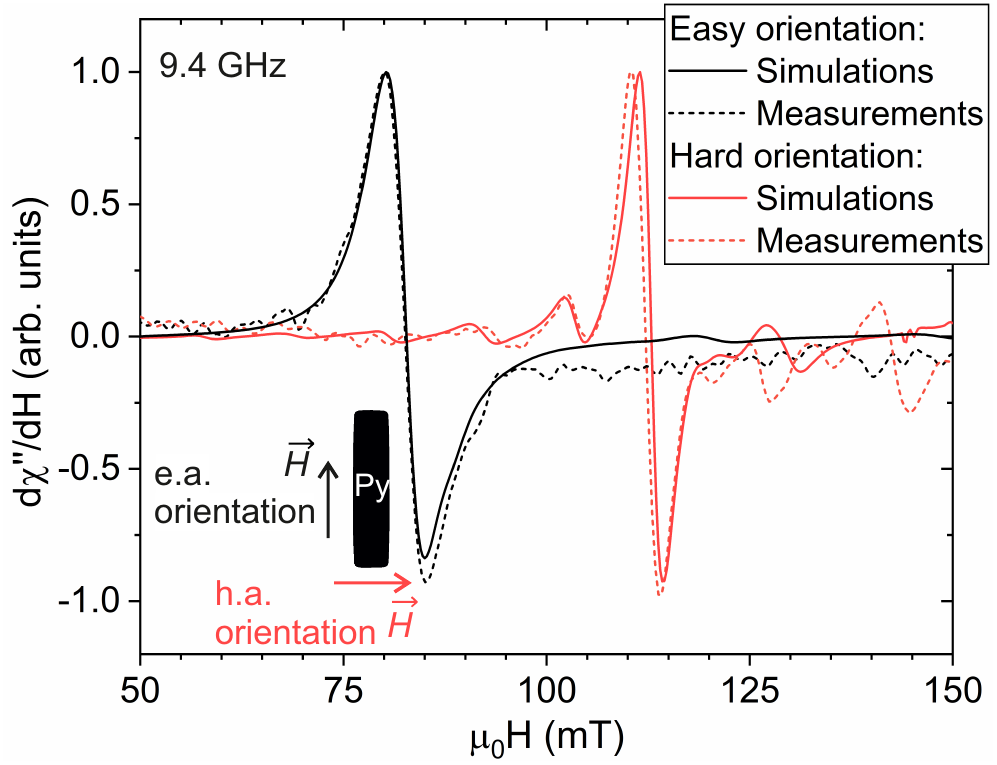}
\caption{Measured and simulated FMR spectra of the slightly rounded rectangular Py microstrip in e.a.\ and h.a.\,orientations. In order to fit the measurements, different damping parameters were used for the simulations in two different field orientations indicated in the plot.}
\label{fig:measfmr}
\end{figure}

The results of the FMR measurements with the use of the multifrequency microantenna [see Fig.\,\ref{fig:antennas}\,(a)] at $f_\mathrm{MW}=9.4$\,GHz are plotted as black and red dashed lines in Fig.\,\ref{fig:measfmr} for the fabricated single Py microstrip in e.a. and h.a.\,orientations, respectively. The nominal dimensions of the strip are the same as for the simulations of the rectangular strip. The FMR spectra in both orientations include the quasi-uniform and spin-wave FMR signals. Additionally, the field gap between the main FMR signal positions in e.a.\, and in h.a.\,orientations is 29.5\,mT for the measurements, which is 10.5\,mT smaller than for the initial simulations shown in Figs.\,\ref{fig:initsims}\,(a,b). A possible reason for these discrepancies could be a deviation of the saturation magnetization from the value used in simulations, which can cause a shift in the resonance field. Furthermore, the quality (roughness) of the edges of the fabricated strip can cause not only the resonance field to shift due to shape anisotropy, but also a change in the effective damping parameter. In addition, the lateral shape of the produced strip is not perfectly rectangular, but has slightly rounded corners and edges [see Fig.\,\ref{fig:stripDem}\,(b)], which can change the relative resonance field positions of all modes. Finally, a deviation from the initial thickness of the strip can be a reason for an altered resonance field, as well.

In order to match the simulated with the measured FMR spectra, the simulations were adjusted taking into account measuring settings and possible effects from the discrepancies described above. As a starting point, the frequency was set to $f_\mathrm{MW}=9.4$\,GHz that was used for the actual FMR measurements. Lowering the frequency shifted all resonances to smaller fields. Further, several simulation parameters were varied and set to new values. As a result, the saturation magnetization was set to $M_\mathrm{s}=730$\,kA/m, compared to $M_\mathrm{s}=700$\,kA/m in Figs.\,\ref{fig:initsims}\,(a,b), which shifted the resonances of both orientations to lower fields and increased the gap between the e.a.\,and h.a.\,main resonances. The damping parameter was set to $\alpha=0.013$ for the e.a.\,and $\alpha=0.008$ for the h.a.\,orientation, compared to $\alpha=0.06$ for both orientations in Fig.\,\ref{fig:initsims}, which broadened the linewidth of all resonances. The reason for the increased damping parameter in e.a.\,orientation was an optimal match to the measured FMR linewidth. However, in the measurements most probably an angle-dependent inhomogeneous broadening is observed rather than a change of the Gilbert damping. The lateral shape of the strip was rounded (see the inset in Fig.\,\ref{fig:measfmr}) to be as close to the SEM image of the sample and its thickness was set to 25\,nm, which decreased the gap between the e.a.\,and h.a.\,resonances and slightly the gaps between the resonances within one orientation, respectively. The cell size was kept the same as for the rectangular strip simulations. In Fig.\,\ref{fig:measfmr} the simulated FMR spectra using adjusted parameters for the strip in e.a.\,and h.a.\,orientations are plotted as black and red solid lines, respectively. The adjusted simulations show a good match to the measured spectra for the main FMR line and spin-wave signals. The spin-wave signals are much less visible in the simulated FMR spectrum using adjusted parameters for the strip in e.a.\,orientation compared to the initial simulations plotted in Fig.\,\ref{fig:initsims}\,(a). The reason for that can be an increased damping in the sample as the increased linewidth causes an overlap of the quasi-uniform and the spin-wave excitations. Another reason can be the change in shape anisotropy due to the rounded edges of the strip in the adjusted simulations, which decreases the resonance field difference between the spin waves.

In the FMR spectrum measured in h.a.\,orientation several spin-wave signals below and above the main line are recognizable (see red dashed line in Fig.\,\ref{fig:measfmr}). The signals below the main FMR line fit the simulations nicely, whereas the signals above show some deviation. A possible reason for that is the quality of the edges, i.e.\, the presence of defects etc., which shifts the resonance field of the localized modes to lower values \cite{044}. However, edge quality is not taken into account in the adjusted simulations apart from the adapted damping parameter. The influence of the edge quality is more noticeable when the spin wave is localized closer to it. As can be seen from Fig.\,\ref{fig:initsims}\,(f) spin waves observed above the quasi-uniform excitation are formed along the longer edges of the strip. Thus, the discrepancy is stronger for these spin waves and as a result their FMR signals. The adjusted simulations confirmed by the FMR measurements show that a variety of spin waves can be excited at 9.4\,GHz in the single Py strip with nonperfect rectangular shape. Spin waves can be excited in both investigated orientations, e.a. and h.a. Due to the shape anisotropy and increased damping parameter in e.a.\,orientation spin waves are less pronounced compared to the h.a.\,orientation.

\begin{figure*}
\includegraphics[width=0.8\textwidth]{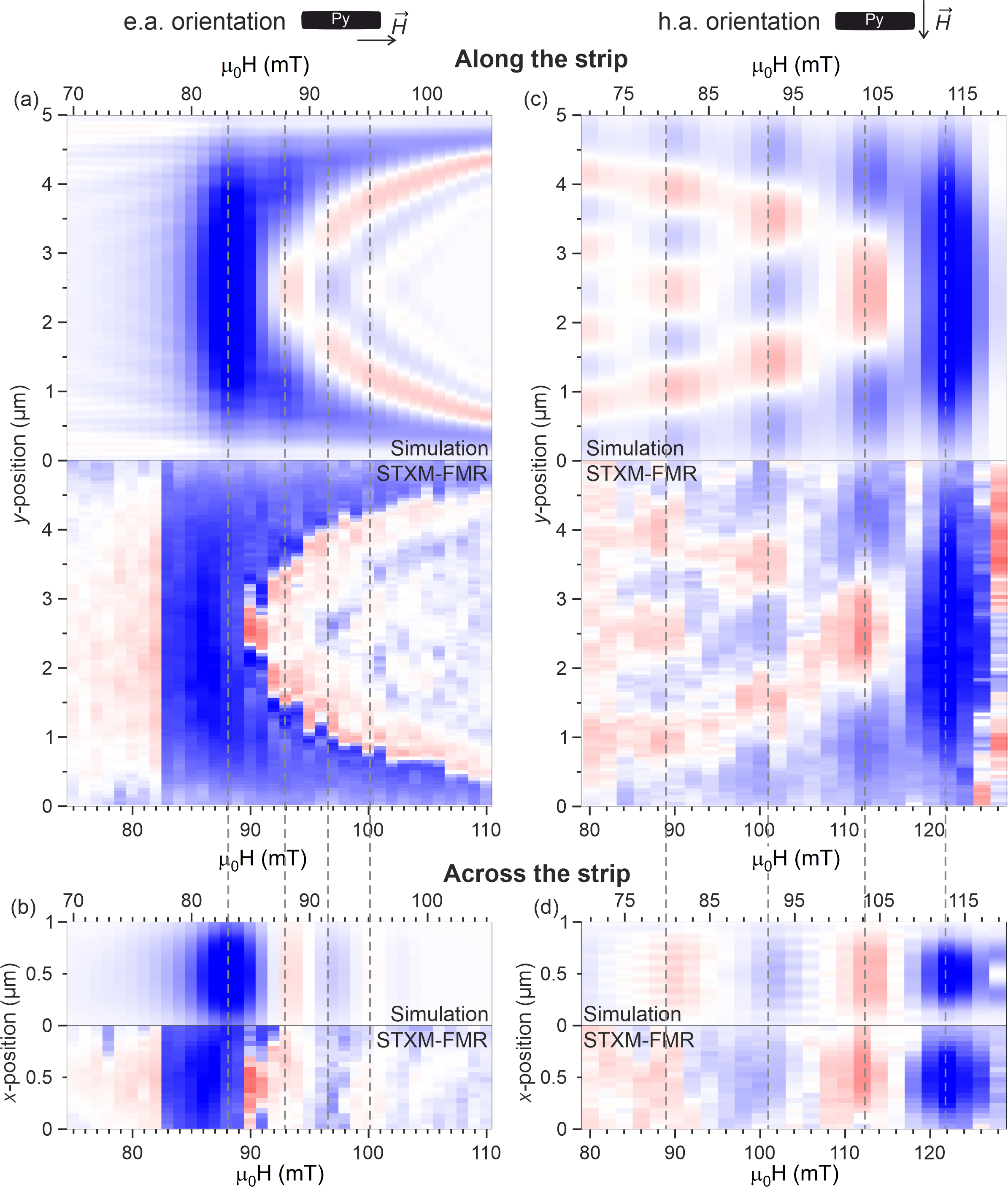}
\caption{Overview of the simulated spin-wave profiles and the combined and normalized amplitude\slash phase data extracted from the STXM-FMR measurements: (a,c) along and (b,d) across the Py strip in (a,b) e.a.\ and (c,d) h.a.\,orientations.}
\label{fig:stxmfmrstat}
\end{figure*}

\subsection{\label{sec:stxmfmr}STXM-FMR Imaging}
The measured STXM-FMR data consists of seven images representing seven equidistant time-steps of the the $m_\mathrm{z}(t)$ distribution across the sample depicting dynamics over one excitation and, thus, magnetization precession cycle \cite{038,028}. By using a temporal discrete Fourier transform at each point in the scan the amplitude and the phase at the MW frequency was extracted from the measured dynamics \cite{028,129}. Via a similar approach as for the simulations (see Fig.\,\ref{fig:initsims}), the amplitude and the phase data extracted from the measurements at several external field values was combined into profile overview amplitude and phase sets (not shown here). In order to be able to compare those overview sets of the STXM-FMR results with the simulated spin-wave profiles, another step of data processing was performed: the amplitude and the phase overview data were combined into one plot by multiplying the amplitude data with the sine of the phase data and normalizing the result (to the [-1;1] range).

\begin{figure*}
\includegraphics[width=0.65\textwidth]{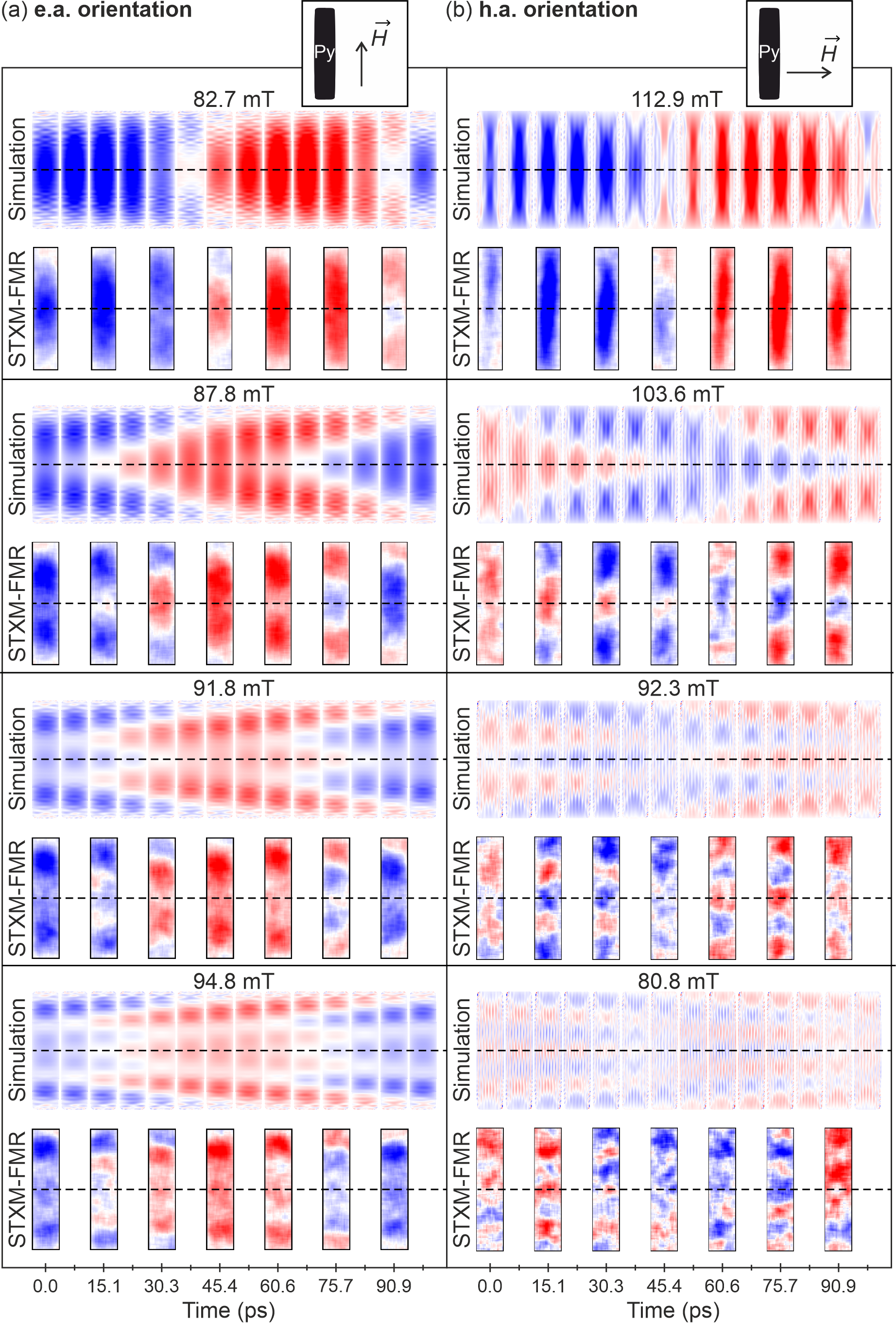}
\caption{Simulated and STXM-FMR-measured time evolution of $m_\mathrm{z}(t)$ over one period for  the entire strip in the (a) e.a.\ and (b) h.a.\,orientations at four different resonance field values for each orientation.}
\label{fig:stxmfmrdyn}
\end{figure*}

The resulting overview plots of the STXM-FMR data and the simulations along (a,c) and across (b,d) the strip in e.a.\,(a,b) and in h.a.\,(c,d) orientations, respectively, are shown in Fig.\,\ref{fig:stxmfmrstat}. The vertical gray dashed lines indicate the correlation between the measurements and the simulations at resonances. For comparison the adjusted simulations (see Fig.\,\ref{fig:stxmfmrstat}) are used with the damping parameter $\alpha=0.008$ for both orientations. This value of the damping parameter was used for the h.a.\,orientation of the strip in adjusted simulations, as it gives a good match between the simulations and the STXM-FMR measurements in both orientations. The latter confirms this way the previous assumption that the line broadening observed in the FMR measurements in e.a.\,orientation was due to angle-dependent inhomogeneous broadening rather than a change of Gilbert damping. In Fig.\,\ref{fig:stxmfmrstat} for better visual matching in the figure, the simulations are plotted with the same field step size as for the measurements, i.e.\ 1\,mT in e.a.\,orientation and 2\,mT in h.a.\,orientation, respectively.

In Fig.\,\ref{fig:stxmfmrstat} by comparing the main FMR signal positions (in both orientations) between the simulations and the combined and normalized amplitude\slash phase data extracted from STXM-FMR measurements, a field offset is observed. This field offset can be a result of the field calibration error of the measurement setup, and\slash or a possible saturation magnetization difference between the samples fabricated for the FMR and STXM-FMR measurements. In e.a.\,orientation the offset is approximately 5\,mT and in the h.a.\,orientation it is 9\,mT. The reason for the different offset in the two orientations is that the change of sample orientation involves remounting and reconnecting the sample with the microantenna. Another evidence of the offset not being a result of the sample shape difference between the measurements and the simulations is the relative position of the resonances in each orientation. As can be seen from the initial (see Fig.\,\ref{fig:initsims}) versus adjusted (see Fig.\,\ref{fig:measfmr}) simulations, the change in the sample shape also changes the relative positions of the resonance lines within one orientation. In Fig.\,\ref{fig:stxmfmrstat} the field gaps between the resonances measured with STXM-FMR are the same as in the simulations, meaning that the shape and the thickness were simulated very closely to the real sample. Hereafter, spin-wave profiles corresponding to the FMR lines will be referred to using the field values from the simulations. The overall comparison of the spin-wave overviews in the range of fields in both orientations of the strip along and across the strip in Fig.\,\ref{fig:stxmfmrstat} reveals a good agreement between the simulations and the STXM-FMR measurements. In the h.a.\,orientation clear spin-wave patterns can be observed along and across the strip at different static external field values [see Figs.\,\ref{fig:stxmfmrstat}\,(c,d)]. Additionally, the spin-wave profile overview plot of the h.a.\,orientation shows a clear separation between the spin waves [Fig.\,\ref{fig:stxmfmrstat}\,(c)] in contrast to the e.a.\,orientation [Fig.\,\ref{fig:stxmfmrstat}\,(a)].

Time-dependent images (snapshots) from simulations and STXM-FMR scans of the spatial distribution of the out-of-plane dynamic magnetization component $m_\mathrm{z}(t)$ within the entire strip in the e.a.\,and the h.a.\,orientations are shown in Figs.\,\ref{fig:stxmfmrdyn}\,(a) and (b), respectively. The quasi-uniform excitation and the three field-adjacent spin waves formed along the longer edge of the strip, which correspond to measured FMR lines above the main resonance in e.a.\,orientation and below the main resonance in h.a.\,orientation, are shown in the figure. The time-series of images are stacked in rows. The STXM-FMR scans in a row depict 7 measured snapshots of one magnetization precession period, while the simulations represent 14 points of the same period. The simulated images fit well the measured spin-wave configuration and dynamics over the entire precession period. Both, measurements and simulations, unexpectedly reveal a nonstanding character of the spin waves at resonance in a single confined microstrip. That can be concluded from tracking the position of the amplitude minima of the waves, which in the images are the white borders between the red and blue regions. For example, in the e.a.\,orientation at 87.8\,mT [see Fig.\,\ref{fig:stxmfmrdyn}\,(a)] one can see that the amplitude minima's positions move from the center of the strip to the shorter edges during half of a precession period. The same is true for the h.a.\,orientation at 103.6\,mT [see Fig.\,\ref{fig:stxmfmrdyn}\,(b)], but in the opposite direction, the amplitude minima change their position from the shorter edges of the strip to the center within half a period. Similar spin-wave behavior was observed in the simulations of the perfectly rectangular strip in both orientations at the fields corresponding to spin waves similar to the shown ones in Fig.\,\ref{fig:stxmfmrdyn}. The observed nonstanding character of the spin waves shown in Fig.\,\ref{fig:stxmfmrdyn} is most probably a result of the inhomogeneity of the internal field as was shown in \cite{019}, i.e. its gradual decrease closer to the edges of the strip along the direction of an external static magnetic field. The same nonstanding character is observed at other resonance fields shown in Fig.\,\ref{fig:stxmfmrdyn}, at 91.8\,mT and 94.8\,mT in e.a.\,orientation and 92.3\,mT and 80.8\,mT in h.a.\,orientation, respectively. In general the STXM-FMR measurements show a good agreement with the adjusted simulations both, in the snapshot spin-wave analysis using overview plots and in the time-dependent spin-wave behavior.

\section{\label{sec:disConcl}Discussion and Conclusion}
It was shown before that in a rectangular microstrip excitation of a variety of resonances, from quasi-uniform excitation to spin waves with a uniform periodic field, is possible \cite{025,026,038}. Moreover, when an external field is applied in the plane, a strong demagnetizing field gradient causes a magnetization inhomogeneity throughout the strip area and a rotation of the magnetic orientation closer to the edges of the strip, which are perpendicular to the external field \cite{019}. Consequently, the effective field in a rectangular microstrip is highly inhomogeneous \cite{057}. As a result, the spin-wave dispersion also changes from the center of the strip towards its edges \cite{046}. Previously reported measurements using time-resolved Kerr microscopy had a limited spatial resolution, insufficient to observe the time evolution of more localized spin waves \cite{056}. In our study STXM-FMR measurements in combination with micromagnetic simulations allow us to investigate the time-evolution of spin waves over a static magnetic field range within a single thin rectangular Py microstrip. The experimental results are in good agreement with micromagnetic simulations. Somewhat unexpectedly, the results reveal a nonstanding character of the excited spin waves in the single confined microstrip at and off resonance, the latter being detected with microantenna-based FMR as well. A non-standing character was originally not expected, because only a uniform excitation field is applied to the specimen during the measurements, suggesting standing spin waves \cite{034}, and also no additional magnetic microstructures are located in close vicinity of the strip, as opposed to the sample system in \cite{038}.

A reason for the non-standing spin-wave character could be the effective field gradients in regions closer to the edges of the strip, which act as centers of inhomogeneous excitation of the waves. When analyzing the spin-wave excitations within the confined microstructure, it should be taken into account that the eigenmodes' dispersion and, thus, the resulting spin waves depend strongly on the internal field distribution \cite{008,019,048}. The internal field within the microstrip is highly inhomogeneous due to its confined size, especially in the direction parallel to the external static magnetic field. Therefore, the spin-wave dispersion varies along this direction, i.e., spin-wave eigenmodes change their wavelength\slash phase during propagation. Another reason could be related to the propagation length of the spin waves, which can be smaller than the confinement size of the strip, meaning that the spin waves do not\slash  only insufficiently reach the opposite edge of the strip and, thus, do not\slash only insufficiently get reflected to form a standing wave.

Moreover, the influence of the internal field distribution on the spin-wave behavior was observed, when the lateral shape of the strip was changed from a perfect rectangle to a rectangle with rounded corners and edges, and, additionally, when the thickness of the strip was decreased in the adjusted micromagnetic simulations, exploiting that the internal field distribution depends on the overall shape of the strip. That demonstrates that by changing the overall shape of the strip it is possible to shift the mutual arrangement of the spin waves in field at the same MW frequency. The latter allows to modify, for example, the spin configuration of some particular spin waves. Our findings provide important insights into the spin-wave dynamics in rectangular confined microstructures and their evolution during the magnetization precession period.

\begin{acknowledgments}
We thank the HZB for the allocation of synchrotron radiation beamtime. We also thank T. Feggeler, H. Stoll, and J. Gr\"afe for their help during the STXM-FMR measurements and A. Halilovic for her valuable contribution to the lithography process. Additionally, we would like to thank M. Bechtel for technical support at the beamline. The authors would like to acknowledge financial support by the Austrian Science Foundation (FWF) via project No I-3050. S. Mayr would like to acknowledge funding from the Swiss National Science Foundation under Grant No. 172517.
\end{acknowledgments}

\bibliography{slReferences}

\end{document}